\documentclass[bbm,amsfonts,amsmath,amssymb,12pt]{article} 
\usepackage[latin1]{inputenc} 
\usepackage[T1]{fontenc} 
\usepackage[english]{babel} 
\usepackage{amssymb,amsmath} 
\usepackage{amscd} 
\usepackage{fancyhdr} 
\usepackage{color} 
\usepackage{hyperref} 
 
\usepackage{epsfig} 
\usepackage{graphicx}
\usepackage{bm}
\usepackage{subfigure} 
 
\definecolor{red}{rgb}{1., 0., 0.} 
\definecolor{blue}{rgb}{0., 0., 1.} 
\definecolor{green}{rgb}{0.1, 0.7, 0.} 
\definecolor{purp}{rgb}{1,0,1}

\newcommand{\bc}{\begin{cases}\begin{aligned}} 
\newcommand{\ec}{\end{aligned}\end{cases}}

\newcommand{\eq}{\begin{equation}} 
\newcommand{\fine}{\end{equation}}

\begin{document}  
\begin{titlepage}  
\vspace {-3cm} 
 \begin{center}
\large \bf Higher  Representations: Confinement and Large N 
 \end{center}
\vskip .5cm \centerline{Francesco {\sc Sannino}\footnote{E-mail: sannino@nbi.dk}} \vskip 0.3cm 
\begin{center} 
\small The Niels Bohr Institute, Blegdamsvej 17, DK-2100 Copenhagen \O, Denmark
\end{center} 
\vskip .4 cm 

\begin{abstract} 
We investigate the confining phase transition as function of temperature for theories with dynamical fermions in the two index symmetric and antisymmetric representation of the gauge group. By studying the properties of the center of the gauge group we predict for an even number of colors a confining phase transition, if second order, to be in the universality class of Ising in three dimensions. This is due to the fact that the center group symmetry does not break completely for an even number of colors.  {}For an odd number of colors the center group symmetry breaks completely. This pattern remains unaltered at large number of colors. We claim that the confining/deconfining phase transition in these theories at large N is not mapped in the one of super Yang-Mills. The phase transition in super 
Yang-Mills is expected to be first order since the center of the gauge group remains intact for any number of colors. 

We extend the Polyakov loop effective theory to describe the confining phase transition of the theories 
studied here for a generic number of colors.

Our results are not modified when adding matter in the same higher dimensional representation of the gauge group. We comment on the 
interplay between confinement and chiral symmetry in these theories and suggest that they are ideal laboratories to shed light on 
this issue also for ordinary QCD. 

We compare the free energy as function of temperature for different theories. We find that the conjectured thermal inequality between the infrared and ultraviolet degrees of freedom computed using the free energy does not lead  
to new constraints on asymptotically free theories with fermions in higher dimensional representation of the gauge group. 

 Since the center of the gauge group is an important quantity for the confinement properties 
at zero temperature our results are relevant here as well.
\end{abstract}

\end{titlepage} 
\section{Introduction} 
Understanding strong dynamics is a challenging and fascinating 
problem. Any progress in this area is likely to affect our 
understanding of Nature. 

We investigate the confining phase transition as function of temperature in theories with fermionic matter in higher dimensional representation of the gauge group. We are interested, specifically, in theories with dynamical fermions in the two index symmetric or antisymmetric representation of the gauge group (denoted here A/S-theories). 

The one flavor case is interesting due to the recent observation made in \cite{ASV} that the A/S theories share the nonperturbative bosonic sector with the one of super Yang-Mills at 
a large number of color. This limit was introduced by Corrigan and Ramond (CR) \cite{Corrigan:1979xf} as an alternative to the 't Hooft large N limit of one-flavor QCD \cite{'tHooft:1973jz}. To this end CR considered a Dirac fermion in the two index antisymmetric representation of the gauge group. However when referring here to the CR limit we consider also the symmetric representation as done in \cite{ASV}. Note that the one flavor QCD with three colors can also be understood as a theory with a Dirac fermion in the two index antisymmetric representation. In the CR large N limit of QCD
the fermions loops are not suppressed and hence the properties related to the axial anomaly can be captured better than in the standard 't Hooft limit. 
Some of the large N  consequences for the CR limit were investigated by Kiritsis and Papavassiliou \cite{Kiritsis:1989ge}. A partial identification of a sector of these theories with super Yang-Mills was only proposed recently by Armoni, Shifman and Veneziano in \cite{ASV}. The link with supersymmetry \cite{ASV} seems to allow for a better control over a number of observables. Such a relation has also been used in \cite{Feo:2004mr} to export part of the phenomenological knowledge about QCD to make predictions on the low energy spectrum of super Yang-Mills. 

The phase diagram as function of number of A/S flavors versus the number of colors has been studied carefully in \cite{Sannino:2004qp}. To extract further quantitative results, at finite number of colors, the effective Lagrangian, of the Veneziano-Yankielowicz \cite{Veneziano:1982ah} type\footnote{The Veneziano-Yiankielowicz effective Lagrangian for super Yang-Mills needs to be extended \cite{Merlatti:2004df} to be able to describe other relevant properties of super Yang-Mills such as domain walls \cite{Merlatti:2005sd} or the glueball sector of the theory.}, for the one flavor A/S-type theories was constructed in \cite{Sannino:2003xe}. 

Recently it has also been shown that theories with just two flavors of S-type matter are natural candidates for breaking the electroweak symmetry dynamically while also yielding a very light composite Higgs \cite{Sannino:2004qp,Hong:2004td,Dietrich:2005jn}. These new technicolor theories are not ruled out by electroweak precision data.

Here we analyze the confining properties of a generic A/S theory as function of temperature and compare them with the ones expected in (super) Yang Mills and theories with matter in the fundamental representation. We will demonstrate that even at infinite number of colors the confinement properties cannot be mapped into super Yang-Mills. This can be explained using both; the formal language of the Wilson lines extending in the temporal direction (i.e. the Polyakov loop) as well as the free energy language which although related to the Wilson line argument may be more intuitively understood. 
While the two index representation is not in the universality class of super Yang-Mills when discussing confinement properties at any number of colors the A and S theories do display the same properties. This means that there is still a limit in which the two different theories are completely equivalent in the CR large N limit but they do not share the confining properties of super Yang-Mills. This is not the case in the 't Hooft limit in which the large N theory matches the confinement properties of Yang-Mills. Note that at nonzero temperature the confining properties of the theory are investigated using the center symmetry of the gauge group which is independent of supersymmetry and it is a global property of the theory. 
In order to make our results more transparent we construct the simplest Polyakov loop effective potentials following Svetitsky and Yaffe \cite{Svetitsky} and more recently Pisarski and collaborators \cite{Pisarski}. 

We study the free energy which helps building our intuition about the physical behavior of different gauge theories and 
 the different large N limits. We will study the different thermal degrees of freedom (d.o.f.) which are active at different temperatures. At very high and low temperatures for an asymptotically free theory which confines in the infrared (or Higgses as for chiral gauge theories) the number of d.o.f. linked to the free energy of the theory can be computed exactly. We show, perhaps not surprisingly, that in the CR limit the ultraviolet d.o.f. match the one of super Yang-Mills for the one flavor A/S-type theory and in the infrared for any number of colors there is a mass gap and hence no light d.o.f.. In the 't Hooft limit of QCD one recovers the pure Yang-Mills d.o.f. in the ultraviolet but not in the infrared. This is so since at large N \`{a} la 't Hooft one expects the axial symmetry to break spontaneously  yielding a Goldstone boson. One can say that the CR limit better describes the properties of one flavor QCD which does not display such a massless boson. Note that we did not use any explicit property of super Yang-Mills in these two extreme temperature regimes, however given that in these two extreme temperature limits the thermal d.o.f. match the super Yang-Mills one, we may be tempted to conjecture that the A/S theories at large N also display a behavior similar to super Yang-Mills at the confining phase transition. However this is not the case, since the center group symmetries are very different even at infinite number of colors, i.e. the full center group breaks always in the A/S theories while it does not for super Yang-Mills. 
The center group symmetry is relevant for confinement not only at finite temperature \cite{hep-lat/9610005,hep-lat/9907021}. A/S theories can be studied in string theory (see \cite{hep-th/0412234,hep-th/0407038} for a review). It would be interesting to understand how the different center group symmetry properties are encoded in the string picture. 

When adding more A/S-type flavors the results 
about the order of the confining phase transition as well as large N arguments do not change. We comment then 
on the relation between chiral symmetry and confinement using our recent understanding of the interplay between 
these two phase transitions \cite{DC} in ordinary QCD and alike theories. 

We have also explored the degree of freedom count at finite temperature when varying the number of A/S flavors. 
We find that the conjectured inequality between the ultraviolet and infrared thermal 
d.o.f. introduced by Appelquist, Cohen and Schmaltz (ACS) \cite{Appelquist:1999hr} is still satisfied but it does not provide new constraints on the conformal window of the A/S theory as function of the number of flavors. We remind 
the reader that the ACS conjectured constraints on strongly coupled theories yield results which are in agreement with the conformal windows in supersymmetric asymptotically free theories summarized in \cite{Intriligator:1995au}\footnote{The guiding principle suggested in 
\cite{Appelquist:2000qg} according to which asymptotically free gauge theories choose the infrared realizations minimizing the number of thermal d.o.f. compatible with the 't Hooft anomaly matching conditions together with the Vafa-Witten constraint is unaffected by our results.}. 
In \cite{Appelquist:1999vs} it was also provided an example based on a product of gauge groups first discussed by Georgi \cite{Georgi:1985hf} where 
the inequality does not lead to new constraints.  However, according to the analysis done in \cite{Sannino:2004qp} we expect in the theories 
investigated here a critical 
number of flavors (lower than the value above which one looses asymptotic freedom) below which the theory is expected to confine while above the 
theory develops a strongly interacting infrared fixed point. The latter is expected to persist when increasing 
the number of flavors up to the point above which one looses asymptotic freedom. This is similar to the case with flavor fermions in the fundamental 
representation of the gauge group for which the ACS inequality seems to provide a constraint \cite{Appelquist:1999hr}.

The ACS inequality is one of the possible 4d conjectured inequalities among the infrared and ultraviolet degrees of freedom. There is, indeed, 
widespread hope that there exists a 4d analog of the Zamolodchikov's 2d c-theorem \cite{Zamolodchikov:1986gt}. This means that there exists a {\it 
central charge}, which counts the number of degrees of freedom of a quantum field theory and monotonically decreases along RG flows to the IR, as degrees of freedom are integrated out. Cardy \cite{Cardy} also conjectured that an appropriate 
such central charge may be the coefficient `a' of a certain curvature-squared term, i.e. the Euler density, of the conformal anomaly on a curved space-time background. The conjectured a-theorem is then that the RG flows satisfy $a_{IR}<a_{UV}$.  Cardy's conjecture is supported by recent studies \cite{atheorem}. The a-theorem is particularly relevant for supersymmetric theories (see \cite{atheoremSusy} and references therein) but we will not consider it here.  
 
In the next section we study the formal properties of the $SU(N)$ gauge theory with fermions in the two index representation under the center of the gauge group. We show that the S/A type theories, even at infinite number of colors, and accordingly if the number of colors is even or odd preserve either a partial or no center group symmetry. Since super Yang-Mills preserves the full $Z_N$ center we expect different confinement/deconfinement mechanisms.  We suggest that for an even number of colors the S/A theory, if second order, the confining/deconfining phase transition is in the universality class 
of Ising. We compare our results with QCD and more generally with theories with fermions in the fundamental representation of the gauge group. 
The CR limit of QCD holds for the thermal properties of the theories, even though super Yang-Mills properties cannot be used. 
In section \ref{tre} we construct the Polyakov loop model for the S/A theories while in section \ref{quattro} we discuss the free energy of the 
one flavor theories in different representations and compare them with (super) Yang-Mills. In section \ref{cinque} we add flavors of the S/A type and study the degrees of freedom counted according to the free energy of the theory as function of temperature. We first discuss the interplay between chiral symmetry and confinement and then propose these theories as ideal laboratories to test our current knowledge on the relation 
between chiral symmetry and confinement also for ordinary QCD. We also show that the ACS conjectured inequality does not lead to new constraints on strongly coupled theories with flavor matter in higher dimensional representations of the gauge group. Finally we conclude in section \ref{sei}.
\section{Thermal Wilson Line}
\label{two}

In any $SU(N)$ gauge theory one can define, according to 't Hooft \cite{'tHooft:1977hy,'tHooft:1979uj},  a global $Z_N$ symmetry which naturally emerges from the associated local gauge symmetry. To review the basic properties consider a generic gauge theory $SU(N)$ with a gauge boson and a A/S Dirac fermion. Defining with $\Omega$ the generic gauge transformation in the fundamental representation we have the following transformation laws for the covariant derivative and fermion in a two index representation:
\begin{eqnarray}
D_{\mu} \rightarrow \Omega^{\dagger} D_{\mu} \Omega \ , \qquad q_{c_1,c_2} \rightarrow { \Omega^{\dagger}}_{c_1}^{b_1} {\Omega^{\dagger}}_{c_2}^{b_2} q_{b_1,b_2} \ .
\end{eqnarray}  
We did not differentiate between symmetric and antisymmetric representation of the gauge group. It is also clear that it is straightforward to generalize the transformation rule for a generic irreducible representation of $SU(N)$. 
For example, if rather than considering a fermion 
in the complex representation of the gauge group, one would have chosen a fermion (say $\lambda$) in the adjoint representation of the gauge group (Dirac or Majorana is irrelevant) it would have transformed according to the gauge group as:
\begin{eqnarray}
\lambda \rightarrow \Omega^{\dagger} \lambda \Omega \ .
\end{eqnarray} 
An element of $SU(N)$ must also satisfy the defining properties:
\begin{eqnarray}
\Omega^{\dagger} \Omega = {\mathbf{1}} \ , \qquad {\rm det}\Omega=1 \ .
 \end{eqnarray} 
 Being a gauge transformation $\Omega$ is, in general, space and time dependent. However one can always consider a gauge transformation of the type:
\begin{eqnarray}
\Omega_{c} = e^{i\delta} {\mathbf 1} \ , 
\end{eqnarray}
with $\delta$ a constant phase. Due to the determinant constraint the phase must assume only the following N values:
\begin{eqnarray}
\delta = 2\pi\frac{k}{N} \ , \quad {\rm with} \quad k =0, 1, \ldots \ , N-1 \ .
\end{eqnarray}
This defines a global\footnote{An integer cannot change continuously as function of the space-time.} $Z_N$ symmetry which is the center of the group $SU(N)$. This symmetry assumes a relevant role when considering the equilibrium thermal properties of a generic gauge theory. 

To appreciate how the center group symmetry plays a role at nonzero temperature it is convenient to consider
 the picture in which the temperature $T$ arises as the imaginary time coordinate $\tau$ in the Euclidean space-time. To be more specific the imaginary time is now a compact space 
\begin{eqnarray}
0\leq \tau \leq \beta=1/T \ . 
\end{eqnarray}
Quantum statistics impose that bosons must be periodic in $\tau$ while fermions must be antiperiodic: 
\begin{eqnarray}
A_{\mu}(\vec{x}
, \beta) =A_{\mu}(\vec{x},0) \ , \qquad  f(\vec{x}, \beta) =-f(\vec{x}, 0) \ .
\end{eqnarray}
Where with $f$ we indicate a generic fermion in any representation of the gauge group while $A_{\mu}$ is the ordinary vector potential. Any periodic in $\tau$ gauge transformation respects the previous boundary conditions. However there is an interesting class of gauge transformation pointed out by 't Hooft of the type:
\begin{eqnarray}
\Omega(\vec{x},\beta) = \Omega_c \ , \qquad \Omega(\vec{x},0)=1 \ .
\end{eqnarray}
It is clear now that not all of the fields are invariant under this specific gauge transformation. For example the fermions in the adjoint and the gauge fields are invariant under this transformation while fermion in the complex representation are not. To be more explicit consider the transformations:
\begin{eqnarray}
 \Omega_c^{\dagger} A_{\mu}(\vec{x}
, \beta)\Omega_c&=& A_{\mu}(\vec{x}
, \beta)=  A_{\mu}(\vec{x}
, 0) \ , \\&&\nonumber \\
 \Omega_c^{\dagger} \lambda(\vec{x}
, \beta)\Omega_c&=& \lambda(\vec{x}
, \beta)= - \lambda(\vec{x}
, 0) \ , \\&&\nonumber  \\
\Omega_c^{\dagger }\Omega_c^{\dagger }q(\vec{x}
, \beta)&=& e^{-2\,i\,\delta} q(\vec{x}
, \beta) \ne - q(\vec{x}
, 0)  \quad {\rm unless} \quad \delta = \pi \, n 
\end{eqnarray}
with $n$ integer. To show that the adjoint representation of color is invariant under the previous gauge transformation we have used the fact that the transformation at $\beta$ is just a constant phase which commutes with any $SU(N)$ matrix. However the situation is more interesting for the fermions in the two index theory. We find, using the constraint that $k<N$ that for any N the integer $n$ must be less than two. It is also clear that, for even N, there is always an unbroken $Z_2$ symmetry whose two elements are obtained by fixing the parameter labelling the elements of $Z_N$ to $k=0$ and $k=N/2$. This choice corresponds to $n=0$ or $1$. 

If  fermions are in the fundamental representation then the center group symmetry is broken for general N with no unbroken subgroups. Note that for the two index representation, both symmetric and antisymmetric, the two color theory has unbroken center. This is so since the S theory with two colors corresponds exactly to the adjoint of $SU(2)$ while the A fermion in the $SU(2)$ gauge theory is a gauge singlet and the theory is pure YM. Remarkably, symmetry wise, the center group does not differentiate, for any N, between symmetric and antisymmetric representation of the gauge group while it differentiates between the two index representation and the adjoint one.

This is now a good point to introduce the thermal Wilson line and study its properties when fermions are of 
the A/S type.
\begin{eqnarray}
{\mathbf L}(\vec{x}) = {\mathbf P} \exp\left[i\,g\,\int_0^{\beta}A_0(\vec{x},\tau)\,d\tau\right] \ ,
\end{eqnarray}
where $g$ is the gauge coupling constant, and $A_0$ the vector potential in the time direction while $\mathbf P$ is the path ordering definition. The thermal Wilson line transforms under local $SU(N)$ gauge transformations as:
\begin{eqnarray}
{\mathbf L}(\vec{x}) \rightarrow \Omega^{\dagger}(\vec{x},\beta) {\mathbf L}(\vec{x}) \Omega(\vec{x},0) \ .
\end{eqnarray}
We can now define the Polyakov loop as the trace of the thermal Wilson line:
\begin{eqnarray}
\ell = \frac{1}{N}{\rm Tr}\left[{\mathbf L}\right] \ .
\end{eqnarray} 
This object is locally gauge invariant while under a spatially global $Z_N$ transformation it carries charges one since it transforms according to:
\begin{eqnarray}
\ell \rightarrow e^{-i\delta} \ell \ .
\end{eqnarray}
In a pure (super)gluonic theory (or any asymptotically free theory with matter in the adjoint representation of the gauge group) the Polyakov loop \cite{Polyakov:1978vu} can be used as the order parameter for confinement. This is so since at extremely large temperatures, compared to the intrinsic renormalization invariant scale of the theory $\Lambda$, the coupling constant of the theory vanishes and the thermal average of the loop is:
\begin{eqnarray}
\langle \ell \rangle = \exp\left[\frac{2\, i\pi \,m}{N}\right]\ell_c  \ , \qquad m=0,1,\ldots\,(N-1) \ ,
 \end{eqnarray}
 with $\ell_c$ approaching one as $T/\Lambda \rightarrow \infty$. At low temperatures we have $\ell_c=0$ \cite{'tHooft:1977hy}. One can define a critical temperature $T_c$ above which $\ell_c$ is nonzero and below which $\ell_c$ vanishes. This is the behavior of an order parameter. If the change of $\ell_c$ at $T_c$ is discontinuous one has a first order phase transition while if the change is continuous the transition is of second order or higher.

Applying the general theory of critical phenomena one expects the two color gauge theory with symmetry $Z_2$ to display a second order phase transition in the universality class of Ising while for any other number of colors one 
expects, in general, a first order phase transition. This is simply argued on the possible form of the relevant bosonic effective theory near the phase transition\cite{Svetitsky}. We summarize our findings in table \ref{tabella1}.
 \begin{table}
 \begin{center}
 \begin{tabular}{c||c|c}
   Fermion Representation & Center Symmetry & Phase Transition\\
   \hline
   \hline
   Pure Glue & $Z_N$ & 1st Order (N>2) \\
   Adjoint & $Z_N$ & 1st Order (N>2)\\
   2-index (N even) &$Z_2$& 2nd Order, Ising  \\
   2-index (N-odd) & None& --- \\
   fundamental & None&---\\
 \end{tabular}
 \end{center}
 \caption{Order of the phase transition and expected universality class in case the phase transition 
 is second order. }
 \label{tabella1}
 \end{table}
 
This picture changes dramatically in the presence of matter. We have already shown that for matter in the adjoint 
representation the center remains unbroken. The situation is very different with matter in complex representations.  
It turns out that the two index representation has interesting confining properties. One can naturally divide 
the two index theories in even and odd number of colors, for an arbitrary number of flavors even when taking the large N limit. 

{}For example QCD belongs to the odd number of colors and at large odd-N, even in the CR limit, one does not expect a well defined order parameter for the confinement/deconfinement phase transition. 
We predict for any even N a second order phase transition in the Ising universality class as function of the temperature between the confined and deconfined phase. 

We are now in a position to compare our findings with (super) Yang-Mills. The confining properties of the A/S type theories, while undistinguishable from each other at an infinite number of colors, do not match the exact center group symmetry left intact in super Yang-Mills. This is a relevant difference since physically it means that the physical 
degrees of freedom relevant at the confining phase transition are very different from the ones present, at large N, in super Yang-Mills. In the 't Hooft limit due to the fermionic component suppression at a large number of colors we expect the large N theory to display confining properties identical to the associated pure Yang-Mills theory. 

Our findings do not support the 't Hooft over the CR limit of one flavor QCD. Indeed it 
may still be possible that the CR limit is better suited for QCD. What we can say definitely, though, is that 
one cannot infer information about the confining/deconfining phase transition via
super Yang-Mills. The present results should help shaping or opportunely modify the correspondence between the A/S theories and super Yang-Mills 
at a large number of colors.

\section{The effective theory}
\label{tre}

The effective theory for the thermal Wilson line constructed via a gauge invariant \cite{Svetitsky} average over a 
domain of fixed size, parallels the construction of the Ising model on a spin lattice. There are a number of well established reasons behind the choice of the Polyakov loop as the lowest state effective field relevant for the confining phase transition. Arguments based on this effective theory are valid only in the limit in which the long-range physics is dominated by fluctuations in the Polyakov loop order parameter. This is true, for example, if 
the confining/deconfining phase transition is second order. 

\subsection{Yang-Mills}
Svetitsky and Yaffe suggested that for an $SU(N)$ Yang-Mills theory in $d+1$ space-time dimensions the effective theory describing the fluctuations of the order parameter is a $d$-dimensional $Z_N$ symmetric scalar 
field theory for the Polyakov loop. The simplest effective theory preserving the $Z_N$ symmetry is:
\begin{eqnarray}
V_{YM}=m^2|\ell|^2 + \lambda\, |\ell|^4 + c(\ell^N + {\ell^{\ast}}^N) \ .
\end{eqnarray}
Note that the first two terms are $U(1)$ invariant while the last term is only $Z_N$ invariant\footnote{The term $\ell^N-{\ell^{\ast}}^N$ is also $Z_N$ invariant but would violate charge conjugation which is preserved in 
the theory.} and it is the term which renders the phase transition first order. Another interesting $U(1)$ invariant term is $|\ell|^6$ 
which is a marginal operator in three dimensions but will be omitted in the following. {}For recent reviews on the subject
we refer the reader to \cite{hep-ph/0011193,hep-lat/0301023}. 

For two colors, however, the theory has a $Z_2$ invariance, here $\ell$ is real and the confining phase transition has the same critical exponents of the 3-d Ising model \cite{Engels:1989fz,Engels:1992fs}. The 2+1 $SU(2)$ Yang-Mills theory has also a second order deconfinement phase transition \cite{Teper:1993gp} in the universality class 
of the 2 dimensional Ising model \cite{Engels:1996dz}. The phase transition, as function of temperature, occurs when $m^2$ changes sign and becomes negative when $T>T_c$. Near the phase transition $\lambda$ and $c$ are expected to be constants. 

{}For three colors the cubic term in the action renders the phase transition first order \cite{Svetitsky}. Interestingly also the three dimensional 3-state Potts model \cite{Wu:1982ra,Fukugita:1989cs,Gavai:1988yq} has a first order phase transition. 
The absence of an universal behavior in the 3-dimensional $Z_3$-symmetric models suggests that the deconfinement phase transition is first order in $SU(3)$ Yang-Mills. This expectation has been confirmed in lattice simulations \cite{Celik:1983wz,Kogut:1982rt,Gottlieb:1985ug,Brown:1988qe,Fukugita:1989yb,Alves:1990yq}. The two space dimensional theory displays instead a second order phase transition in the universality class 
of the two dimensional 3-state Potts model \cite{Christensen:1992is}. The four number of colors case is also very interesting but we will not discuss it. {}For $N\geq 5$ the phase transition can be of second or first order 
according to the form of the potential. If second order one would still predict critical exponents in the universality class of the $Z_N$-symmetric chiral clock model which in three dimensions is in the universality class of the $U(1)$-symmetric XY-model. In practice the center group symmetry $Z_N$ is dynamically to a continuous $U(1)$ symmetry. This is consistent with the expectation that the term in $\ell^N$ which breaks the $U(1)$ symmetry to $Z_N$ is irrelevant in three dimensions. Numerical simulations performed for the case of $SU(6)$ and $SU(8)$ Yang-Mills theories \cite{Lucini:2002ku} indicate however a first order phase transition, suggesting, perhaps that all of the 3+1 dimensional Yang-Mills theories with a number of colors larger than three have a first order deconfining phase transition without universal behavior\footnote{It 
may be interesting to observe that recently the dependence on the number of colors of the leading pi pi scattering amplitude in chiral dynamics has been studied. Here we demonstrated the existence of a critical number of colors for and above which the low energy pi pi scattering amplitude computed from the simple sum of the current algebra and vector meson terms is crossing symmetric and unitary at leading order in a truncated and regularized $1/N$ expansion `{a} la 't Hooft. The critical number of colors turns out to be $N=6$ and is insensitive to the explicit breaking of chiral symmetry \cite{Harada:2003em}.}. We note that Pisarski and Tytgat have also suggested a second order phase transition behavior to occur at large N \cite{Pisarski:1997yh}. In the meanwhile new numerical results at a large number of colors have been reported in \cite{Bringoltz:2005rr}.

Deconfinement phase transition in gauge theories with fermions in the adjoint representation of the gauge group have 
 been analyzed (see \cite{Engels:2005te} and references therein) and here one expects a behavior similar to the one of pure Yang-Mills theory.

Another set of interesting Yang-Mills theories which have been explored numerically are the ones with $Sp(2N)$ gauge group\footnote{Note that I am using a convention for labelling the symplectic group in which $Sp(2)=SU(2)$.} in which the center is always $Z_2$ for any number of colors. In reference \cite{hep-lat/0312022} it was shown that for $N$ larger than two these theories do not display an universal behavior for the 3+1 dimensional theory. 

Recently de Forcrand and Jahn \cite{deForcrand:2002vs} have compared, on the lattice, the confinement properties of $SU(N)$ and $SU(N)/Z_N$. More precisely they have studied $SU(2)$ and $SO(3)$. They have argued that confinement may be linked to the nontrivial homotopy group $\pi_1\left[SU(N)/Z_N\right]=Z_N$ which is the same for both $SU(N)$ and $SU(N)/Z_N$ rather than directly to the center group symmetry. Interestingly there are only 3 simple non-Abelian Lie groups for which $\pi_1$ is trivial, i.e. the exceptional groups $G_2$, $F_4$ and 
$E_8$. Hence they are not expected to have a deconfinement phase transition at all. Indeed it has been argued in \cite{Holland:2003jy} that the $G_2$ Yang-Mills theory has a crossover between its low and high temperature regimes.

It is then interesting to ask if the Svetitsky and Yaffe's universality arguments can be applied to $SU(N>2)$ gauge theories with fermions not in the adjoint representation of the gauge group. 

\subsection{Fundamental, Adjoint and A/S representation for matter fields} 
How does the Polyakov loop potential changes when fermions are added to the theory? This problem has been investigated by different authors \cite{MoreDC,DC}.  
Here it is sufficient to say that according to the symmetry analysis performed in \cite{DC} the net effect is to introduce symmetry breaking terms in the effective theory for the Polyakov loop. 

When fermions are added only in the adjoint representation of the gauge group the effective potential has the same form as the one for the corresponding pure Yang-Mills theory but the  coefficients are different. 
If fermions are added in the fundamental representation the full $Z_N$ symmetry is broken and the simplest term one can add to make such a breaking apparent at the effective Lagrangian level is:
\begin{eqnarray}
   (\ell + \ell^{\ast}) \ .
\end{eqnarray}
So the effective Lagrangian for the Polyakov loop when fermions are in the fundamental representation of the gauge group becomes:
\begin{eqnarray}
V_{\rm Fund}=m^2|\ell|^2 + \lambda\, |\ell|^4 + c\,(\ell^N + {\ell^{\ast}}^N) + b\,  (\ell + \ell^{\ast}) \ . 
\end{eqnarray}
A number of observations are in order. Higher dimensional operators spoiling the $Z_N$ symmetry are allowed. More specifically one would expect also $(\ell + \ell^{\ast})^2$ terms which should be included in the low energy effective theory. Besides when fermions are present these operators are naturally coupled to the fermion bilinears. 

What happens for the A/S theory? The situation is rather interesting. For any odd number of flavors the center symmetry is completely broken and the effective theory is similar to the one for quarks in the fundamental representation of the gauge group. However for an even number of colors the $Z_N$ symmetry is broken explicitly to $Z_2$. In this case the term linear in $\ell$ is not allowed and one expects an effective theory of the type:
\begin{eqnarray}
V_{\rm N-even}^{\rm 2-ind} = m^2|\ell|^2 + \lambda\, |\ell|^4 + c\,(\ell^N + {\ell^{\ast}}^N) + d\,  (\ell^2 + {\ell^{\ast}}^2) \ .
\end{eqnarray}
Note that the term linear in $\ell$ is essential when trying to break the $Z_N$ symmetry completely and must disappear when N is even. In this case the first relevant term in an analytical expansion in the Polyakov loop field is quadratic in $\ell$. 
One can imagine a continuous number of colors with fermions in the A/S representation and since for any even N the term linear in $\ell$ vanishes the coefficient in front of this term has a periodic type behavior.  A guess for the effective potential which is valid for any N with S or A-type fermions:
\begin{eqnarray}
V^{\rm 2-ind} =&& m(N)^2|\ell|^2 + \lambda(N)\, |\ell|^4 + c(N)\,(\ell^N + {\ell^{\ast}}^N) + d(N)\,  (\ell^2 + {\ell^{\ast}}^2) 
\nonumber \\ &&+{f}(N)\,\left[\sin\frac{N\pi}{2}\right]\,(\ell + \ell^{\ast}) \ .
\end{eqnarray}
Here $m,\lambda,c,d$, and $f$ are smooth, nonsingular and in general nonzero functions of N. 
Although we expect a periodic behavior as function of the number of colors for the coefficient of the linear term we could also have an even power of the sin function. We also argue that for any N larger or equal to four the term reminiscent of the $Z_N$ symmetry is an irrelevant operator and does not affect the second order character of the phase transition. Besides the leading term in $\ell^2$ which breaks the $Z_N$ symmetry we also expect higher order terms preserving $\ell^2$ such as $\ell^4$ and so on which, in fact, are more relevant at large N than the $\ell^N$ term. Modern lattice simulations can help shedding light on this issue.

As we have already mentioned one can consider at least two different types of large N limits. The one in which the 
fermions are in the fundamental representation and the one in which the fermions are in the 
two index representation of the gauge group. In the first case, i.e. the 't Hooft limit, the center group symmetry 
restores at large N. At the effective Lagrangian level one can say that the terms breaking the $Z_N$ symmetry vanish 
at large N. In particular this limit is interesting for QCD since quenched lattice simulations, for example, are very close to realize the 't Hooft picture \cite{Bringoltz:2005rr}. On the contrary even at large N one never recovers the $Z_N$ symmetry in the CR limit. However this limit is well defined since it just states that the A and S type theories do become exactly equivalent to each other at large N while the confining properties do not match the super Yang-Mills ones.

\section{Free Energy and Thermal Degrees of Freedom}
\label{quattro}
The free energy can be seen as a device to probe and count
the relevant degrees of freedom and has been used recently for phenomenological explorations in \cite{Muller:2005en}. It can be computed, exactly, in two regimes 
of a generic asymptotically free theory: the very hot and the very cold one. 

The zero-temperature theory of interest is
characterized using the quantity $f_{IR}$, related to the free energy by
\begin{equation}  \label{eq:firdef}
f_{IR} \equiv - \lim_{T\to 0} \frac{{\cal F}(T)}{T^4}
\frac{90}{\pi^2} \ ,
\end{equation}
where $T$ is the temperature and ${\cal F}$ is the conventionally
defined free energy per unit volume. The limit is well defined if
the theory has an infrared fixed point. 
For the special case of an
infrared-free theory 
\begin{eqnarray}
f_{IR} = \sharp~~{\rm Real~Bosons}~+ ~\frac{7}{4}~\sharp~~{\rm Weyl-Fermions} \ .
\end{eqnarray}
The corresponding expression in the large $T$ limit is
\begin{equation}  \label{eq:fuvdef}
f_{UV} \equiv - \lim_{T\to \infty} \frac{{\cal F}(T)}{T^4}
\frac{90}{\pi^2}\ .
\end{equation}
This limit is well defined if the theory has an ultraviolet
fixed point. For an asymptotically free theory $f_{UV}$ counts the
underlying ultraviolet d.o.f. in a similar way.

In terms of these quantities, the conjectured inequality \cite{Appelquist:1999hr} for any
asymptotically free theory is
\begin{equation}  \label{eq:ineq}
f_{IR} \le f_{UV}\ .
\end{equation}
This inequality has not been proven but it was
shown to be consistent with known results and then used to derive new
constraints for several strongly coupled, vector-like gauge theories.
It was applied to chiral theories in Ref.~\cite{Appelquist:1999vs}. The
principal focus there was on the possibility of preserving the
global symmetries through the formation of massless composite
fermions.

 For theories developing a mass gap the zero temperature free energy vanishes
and in the deep infrared the free energy does not distinguish between super Yang-Mills, a pure Yang-Mills theory or, for example, one flavor QCD. However when increasing the temperature different theories will respond differently 
 to the temperature change via their active d.o.f. at each given temperature. To be more specific let  us compare (super) Yang-Mills, the one flavor theory with the Dirac fermion in the fundamental representation of the gauge group, and the one flavor A or S theories. All of these theories are expected to develop a mass gap.  This means that for all of these theories $f_{IR}=0$ while the associated ultraviolet number of d.o.f. is reported in table \ref{table2}.
 \begin{table}
\begin{tabular}{c||c|c|c|c}
  &YM & 1-flavor fund. & 1-flavor 2-index & SYM \\
  \hline
  \hline
  &&&&\\
 $f_{UV}$& $2(N^2-1)$ & $2(N^2-1)+\frac{7}{2}N$ &$2(N^2-1)+\frac{7}{4}N(N\pm1)$  & $2(N^2-1)(1+\frac{7}{8})$ \\
\end{tabular}
\caption{Ultraviolet thermal d.o.f. for different gauge theories as function of the number of colors.}
\label{table2}
\end{table}
We can also consider now the large N limit of all of these theories in the UV. As it is expected at a large number of colors the free energy of the one flavor theory with fermions in the fundamental returns the pure gauge result. While the two index representation goes into the supersymmetric limit. 

The situation in the infrared is more subtle. Since in the large N limit of the two index theories the fermion contribution is not suppressed the axial symmetry remains explicitly broken at any N and we expect no Goldstone boson to emerge in the large N limit. However this is not the case in the ordinary 't Hooft limit. So that the $f_{IR}$ would be one in the large N limit of the theory with a fermion in the fundamental representation of the gauge group. The picture should help explaining these limits. 

\begin{figure}[htbp]
\begin{center}
\includegraphics[scale=.5]{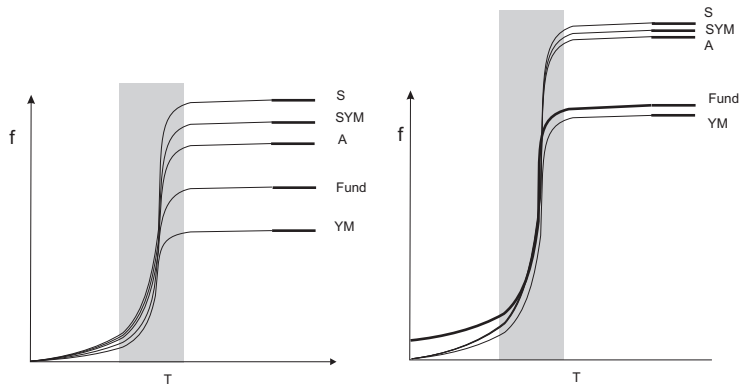}
\end{center}
\caption{Degrees of freedom counted according to the free energy for different theories. 
The left panel represents the degree of freedom count as function of the temperature for a given number of colors. Note that at finite number of colors there are no light degrees of freedom. The right panel represent the large N count of the thermal degrees of freedom. It is clear that the $S/A$ theories become degenerate with SYM both at infinite $T$ and zero $T$. However the degree of freedom count for the fundamental representation disagrees at zero temperature at large N from the YM count since we have a Goldstone boson here.}
\label{free}
\end{figure}

{}We naturally differentiate three regions for the free energy count of freedom as function of temperature: the very hot and cold and an 
intermediate temperature one associate to the confinement transition. The high temperature limit is the one in which one simply counts the ultraviolet d.o.f. of the asymptotically free theories. The two index theories at large N agree with the super Yang-Mills count of degrees of freedom. In the 't Hooft limit 
the theory with a fermion in the fundamental representation overlaps with pure Yang-Mills. At very low temperatures 
we find that since the axial symmetry acting on the fermions (i.e. the $U(1)_R$ for the gluino) is always explicitly broken the free energy count is zero for these theories for any N. However in the 't Hooft limit we expect the emergence of a Goldstone excitation associated to the spontaneously broken axial symmetry at large N. The thermal d.o.f. count at very high and low temperature regions is in agreement with the general expectations  \cite{Corrigan:1979xf,ASV}.  
{}So the CR limit is best suited to describe the low energy temperature regime of one flavor QCD since it does not introduce, at large N, a Goldstone excitation.  

The intermediate temperature regime associated to the confining phase transition is also interesting. 
In the figure we have shaded this region which can also be used t differentiate neatly the 't Hooft from the CR large N limit for one flavor QCD. According to the standard large N limit one predicts a first order phase transition while in the CR limit there is no clear limit at a large number of colors for what concerns the order of the confining phase transition. However if we consider only odd numbers of colors at large N we expect no restoration of the center symmetry. It is however unclear if one can can carry out the limit taking into account also the N-even values of color. Then it is tempting to speculate that due to the oscillatory behavior of the prefactor of the linear term in the effective action of the Polyakov loop one may end up with a near second order phase transition at large N\footnote{As already mentioned above although we expect a periodic behavior, as function of the number of colors, for the coefficient of the term linear 
in the effective potential for the Polyakov loop one can also have an even power of the sin function. In this case no suppression at a large number of colors of the term linear in $\ell$ is expected.}.

\section{Adding Matter}
\label{cinque}

We now add flavors in higher dimensional representations of the gauge group. The previous considerations on the center of the gauge group symmetry do not change. 
\subsection{Confinement versus Chiral Symmetry}
It is interesting now to explore the interplay between chiral symmetry and confinement. This is so since for an odd number of colors we do not have a proper order parameter for confinement while for an even number we do have it and the expected phase transition is second order. Hence the odd number of flavors mimics ordinary QCD while the even number of colors resembles the two color theory with adjoint fermions. However the 
 chiral symmetry pattern can be taken to be the one of QCD for N greater than two for the S-type fermions and different from four for the A-type. Consider, for example, four colors and two S-type flavors.  We expect a second order phase transition both for the flavor and confining phase transition. Chiral symmetry is expected to be restored after the breaking of the $Z_2$ center group symmetry when increasing the temperature\footnote{Here we are assuming that 
the remaining of the full $Z_N$ center group symmetry is still a good order parameter for confinement. If this would not be the case 
than the $Z_2$ phase transition can also occur after chiral symmetry is restored when increasing the temperature.}. Interestingly, according to the findings in \cite{DC} we also predict an induced, quasicritical behavior of the fermionic condensate at the confining phase transition. For an odd number of flavors 
we expect a situation similar to the one in QCD, i.e. the well defined chiral phase transition drives almost critical the behavior of the would 
be order parameter for confinement, i.e. the Polyakov loop \cite{DC}. One would then observe a coincidence in temperature for the exact chiral phase transition and the would be confinement one. Differently than for ordinary QCD, though, even for large number of colors we do not expect the center group symmetry to be restored in these theories. We consider the study of the confining versus chiral symmetry phase transition a relevant and independent test of the present understanding of the relation between confinement and chiral symmetry breaking.

We briefly comment also on the case of $p$ vector like fermions in the fundamental representation and $q$ in the two index representation of the gauge group with $q$ and $p$ such that the theory is asymptotically free. There is no well defined center group symmetry except in the large N limit with even number of colors, where the $Z_2$ symmetry survives.

\subsection{Thermal D.O.F.}

According to ACS $f_{IR} \leq f_{UV}$. However one can imagine different functions of the number of degrees of freedom of an asymptotically free theory which may also satisfy a similar inequality. In four dimensions a well known one is due to Cardy \cite{Cardy}. The hope is 
that inequalities might lead to further constraints on the possible phase structure of a generic strongly coupled gauge theory. Cardy's proposal has been investigated in some detail by Ball and Damgaard in \cite{Ball:2001wr} also for theories with fermions in higher dimensional representations. 
Unfortunately this proposal does not lead to strong constraints. On the other hand the ACS proposed inequality provides interesting constraints on 
the phase diagram of strongly coupled theories. Some of the deduced constraints are validated by known results in supersymmetric  gauge theories \cite{Intriligator:1995au}. Recently we have studied \cite{Sannino:2004qp,Dietrich:2005jn} the phase diagram of theories 
with matter of the A and S type. It is then interesting to see if the ACS inequality can be of use here. 

The d.o.f. count for the multiflavor case is straightforward when considering all of the fermions in the higher dimensional representation of the gauge group and assume, in the infrared, the breaking of the global symmetry group to the maximal diagonal subgroup. Of course we also need the number of flavors to be lower than the number needed to loose asymptotic freedom. 

The ultraviolet number of thermal d.o.f for $F$ Dirac S or A type flavors in a generic $SU(N)$ gauge theory is:
\begin{eqnarray}
f_{UV}=2(N^2 -1 ) + \frac{7}{4}F\,N(N\pm 1) \ .
\end{eqnarray} 
The upper(lower) sign refers to the symmetric(antisymmetric) representation. Assuming the breaking of the $SU(F)_L\times SU(F)_R$  flavor group
to the maximal vector diagonal subgroup $SU(F)$ one can immediately compute the infrared number of d.o.f. provided simply by the number of Goldstone bosons in the theory, i.e.
\begin{eqnarray}
f_{IR} = F^2 - 1 \ .
\end{eqnarray}
This infrared count is incorrect if the number of colors is two for the S-type fermions or if the number of colors is four for the A-type fermions. This is so since in those cases the fermion representations are real (see \cite{Sannino:2004qp} for a detailed discussion of these particular cases).
 According to the one loop coefficient of the beta function asymptotic freedom is lost when:
\begin{eqnarray}
F_{AF}\geq \frac{11}{2}\frac{N}{N\pm 2} \rightarrow \frac{11}{2}\ .
\end{eqnarray}
The upper(lower) sign is for the S(A)-type fermions and in the last step we have assumed the large N limit.
  
Since $f_{IR}$ increases quadratically with the number of Goldstones while the $f_{UV}$ increases linearly with the number of flavors it is sensible to ask what is the number of flavors for which $f_{IR}=f_{UV}$. Here we find:
\begin{eqnarray}
F_c = \frac{1}{8}\left[7N^2 \pm 7N + \sqrt{49N^4 \pm  98N^3 + 177N^2  \pm 64}\right]  \rightarrow \frac{14}{8}N^2\ ,
\end{eqnarray}
where in the last step we have considered the large N limit while the upper(lower) sign is for the S(A)-type fermions. At large 
number of colors we do not differentiate between the S or A type fermions. We find that $F_c$ is larger than $F_{AF}$ and hence the result does not contain any useful information on the phase diagram of these theories as function of number of flavors and colors. It is interesting to 
understand the technical reason which renders the inequality proposed by ACS, as a new constraint on strongly interacting theories, not constraining
here. First, we note that the term in the ultraviolet counting of the thermal fermionic d.o.f. is color enhanced with respect to the case with matter
in the fundamental representation of the gauge group. The infrared number of degrees of freedom is, however, identical to the case with fermions in the fundamental representation. 
It is then clear that one needs a substantially larger number of flavors, with respect to the fundamental representation, for the infrared and the ultraviolet number of degrees of freedom to have the same numerical value. Besides the theory looses asymptotic freedom earlier as function of the number of flavors since we have more screening matter than in the fundamental representation. 

{}From the arguments above it is easy to convince oneself that one can construct other asymptotically free gauge theories in which the inequality between the thermal d.o.f. in the ultraviolet and in the infrared does not lead to new constraints on the strongly coupled theory under investigation. As another example consider a $SU(N)$ gauge theory with $F$ Weyl fermions in the adjoint representation of the gauge group. Here the global symmetry is $SU(F)$ which breaks spontaneously to $O(F)$. In the ultraviolet:
\begin{eqnarray}
f_{UV} = 2(N^2 - 1) +\frac{7}{4}F(N^2-1) = 2(N^2-1)(1+\frac{7}{8}\,F)  \ ,
\end{eqnarray}  
while in the infrared:
\begin{eqnarray}
f_{IR} =  \frac{F^2}{2} + \frac{F}{2} -1 \ .
\end{eqnarray}
Here the number of flavors for which $f_{IR}=f_{UV}$ is
\begin{eqnarray}
F_c =\frac{1}{4} \left[7N^2 -9 + \sqrt{49N^4 - 62N^2 +49}\right] \rightarrow \frac{14}{4}N^2 \ . 
\end{eqnarray}
For any number of colors asymptotic freedom is lost for 
\begin{eqnarray}
F_{AF}\geq \frac{11}{2} \ .
\end{eqnarray}
The thermal inequality does not provide a constraint here as well. The attentive reader has already noticed that the UV properties, i.e. the beta function and the thermal count of the degrees of freedom is identical to the one of  $F/2$ fermions in the two index representation of the gauge group. This observation has been made in \cite{ASV}. However, in the infrared these theories are not equivalent or at most they share certain subsectors \cite{ASV}. Besides the confinement properties of these theories, as for the single flavor   case, are very different in the large number of color limit.

We note that in the supersymmetric or chiral theories investigated in \cite{Appelquist:1999hr,Appelquist:1999vs} one dealt with at most one fermion in the adjoint or two index representation of the gauge group while the inequality set constraints on the number of flavors in the fundamental representation of the gauge group. 
In \cite{Appelquist:1999vs} an example based on a product of gauge groups first discussed by Georgi
\cite{Georgi:1985hf} was investigated  where 
the inequality does not lead to new constraints.  According to the analysis done in \cite{Sannino:2004qp}, however, we expect for the theories 
investigated here a critical 
number of flavors below which the theory is expected to confine while above it the 
theory develops a strongly interacting infrared fixed point. This nontrivial and interacting fixed point is expected to persist when increasing 
the number of flavors up to the number of flavors above which one looses asymptotic freedom. This is similar to the case with flavor fermions in the fundamental 
representation of the gauge group for which the ACS inequality provides a constraint though \cite{Appelquist:1999hr}.

\section{Conclusions and Outlook}
\label{sei}

We have studied the confining phase transition as function of temperature for theories with dynamical fermions in the two index symmetric and antisymmetric representation of the gauge group.  Using the properties of the center of the gauge group we have predicted for an even number of colors a second order confining phase transition in the universality class of Ising.   {}For an odd number of colors the center group symmetry breaks completely. This patterns remains unaltered at a large number of colors. We claim that the confining/deconfining phase transition in these theories at large N is not mapped in the one of super Yang-Mills. In particular we expect the phase transition in super 
Yang-Mills to be first order since the center of the gauge group remains intact for any number of colors. 

We have then generalized the Polyakov loop effective theory to describe the confining phase transition with 
fermions in the S/A type theories. Adding matter in the same higher dimensional representation of the gauge group \
does not alter our conclusions. 
We have also suggested how the confinement and chiral symmetry phase transition may be linked in these theories. We have proposed these theories as ideal laboratories to shed light on 
the same issue for ordinary QCD. 

We compared the free energy as function of temperature for various theories. At very high and low temperatures we can compute it exactly and extract the relevant number of degrees of freedom while at intermediate temperature near the confining phase transition we used the knowledge about the center group symmetry and universality arguments. Since the theory with a fermion in the two index antisymmetric representation of the gauge group for three colors is one flavor QCD we compared the CR large N limit with the 't Hooft one.

Among other things we have shown that the conjectured inequality between the infrared and ultraviolet degrees of freedom computed using the free energy does not lead  
to strong constraints on asymptotically free theories with fermions in higher dimensional representation of the gauge group. This is so since 
the flavor-dependent term of the ultraviolet count of the number of degrees of freedom is enhanced 
with respect to matter in the fundamental representation, while the infrared count is not.

Our findings do not support the 't Hooft over the CR limit of one flavor QCD. Indeed it 
may still be possible that the CR limit is better suited for QCD. What we can say definitely, though, is that 
one cannot infer information about the confining/deconfining phase transition using super Yang-Mills properties. Since the center of the gauge group is a relevant quantity also for the confinement properties 
at zero temperature we expect our results to lead to new relevant information also at zero temperature. The present results help shaping and opportunely modify the correspondence between the A/S theories and super Yang-Mills at a large number of colors. 

\newpage
\centerline{\bf Acknowledgments}
\vskip .5cm
I am very happy to thank S. Catterall,  P.H. Damgaard, L. Del Debbio, D.D. Dietrich, P. Di Vecchia, P. de Forcrand,   M. Frandsen, L. Giusti, J. Greensite,
 P. Merlatti, T.A.~Ryttov,  J. Schechter, M.~Shifman, K. Tuominen and G. Veneziano for helpful discussions, comments or careful reading of the manuscript. 
 
\noindent
The work is supported by the Marie Curie Excellence Grant as team leader under contract MEXT-CT-2004-013510 and by the Danish Research Agency. We also thank the CERN theory division for the kind hospitality during the initial stages
 of this work.

\end{document}